\documentclass[
 reprint,
 amsmath, amssymb,
 aps,
prb,
]{revtex4-2}

\usepackage{graphicx}% Include figure files
\usepackage{dcolumn}% Align table columns on decimal point
\usepackage{bm}% bold math
\usepackage{multirow}
\usepackage{braket}
\usepackage{color}
\usepackage{ulem}
\usepackage{here}
\usepackage{times}
\usepackage{booktabs}

\begin{document}

\title{Rotation and electric-field responses and absolute enantioselection in chiral crystals}

\author{Rikuto Oiwa$^{1}$}
\author{Hiroaki Kusunose$^{1}$}
\affiliation{$^{1}$Department of Physics, Meiji University, Kawasaki 214-8571, Japan}

\begin{abstract}
  Microscopic origin of chirality and possible electric-field induced rotation and rotation-field induced electric polarization are investigated.
  By building up a realistic tight-binding model for elemental Te crystal in terms of symmetry-adopted basis, we identify the microscopic origin of the chirality and essential couplings among polar and axial vectors with the same time-reversal properties.
  Based on this microscopic model, we elucidate quantitatively that the inter-band process, driven by the nearest-neighbor spin-dependent imaginary hopping, is the key factor in the electric-field induced rotation and its inverse response.
  From the symmetry point of view, these couplings are characteristic common to any chiral material, leading to a possible experimental approach to achieve absolute enantioselection by simultaneously applied electric and rotation fields, or magnetic field and electric current, and so on, as a conjugate field of the chirality.
\end{abstract}

\maketitle

%%%%%%%%%%%%%%%%%%%%%%%%%%%%%%%%%%%%%%%%%%%%%%%%%%%%%
{\it Introduction.---}
\label{sec:intro}
Chirality is three-dimensional geometric property exhibiting ubiquitously in nature.
Handedness or enantiomer in chiral materials is characterized by a quantity having time-reversal $\mathcal{T}$-even pseudoscalar (spatial-inversion $\mathcal{P}$ odd) property~\cite{barron_2004,barron_2013}, whose sign corresponds to the left and right handednesses.
This significant quantity in chiral materials, however, has not fully been understood at microscopic level.
Thus, clarifying the microscopic origin of a $\mathcal{T}$-even pseudoscalar inherent in chiral materials is essential to unveil the heart of chirality and to achieve absolute enantioselection in chiral materials.

In this Letter, we begin with a phenomenological discussion of the features of $\mathcal{T}$-even pseudoscalar and important coupling to it based on the symmetry argument.
Then, to confirm the existence of the coupling, we quantitatively investigate the expected responses by using the specific microscopic model of the elemental Te crystal.
Lastly, we propose a possible experimental approach to realize absolute enantioselection in chiral materials, by the external fields that is accessible to $\mathcal{T}$-even pseudoscalar via the elucidated coupling.

{\it Chirality and related quantities.---}
From the symmetry point of view, a $\mathcal{T}$-even pseudoscalar can be decomposed into the inner product of polar and axial vectors with the same $\mathcal{T}$ property.
Such a decomposition is clearly carried out by the concept of electronic multipole basis~\cite{SH_HK_2018, SH_MY_YY_HK_mul_2018, HW_YY_Mul_2018, HK_RO_SH_comp_mul_2020}.
Namely, a $\mathcal{T}$-even pseudoscalar corresponds to an electric toroidal (ET) monopole $G_{0}$ with $(\mathcal{P}, \mathcal{T}) = (-,+)$, and is decomposed into the inner product of a magnetic (M) dipole $\bm{M}$, $(\mathcal{P},\mathcal{T})=(+,-)$, and magnetic toroidal (MT) dipole $\bm{T}$, ($\mathcal{P},\mathcal{T})=(-,-)$, as
\begin{align}
  G_{0} \,\,\,\to\,\,\, \bm{T} \cdot \bm{M}.
  \label{eq:G0_TM}
\end{align}
Since a conjugate field of $\bm{T}$ is an electric current $\bm{J}$, Eq.~(\ref{eq:G0_TM}) indicates that linear current-induced magnetic response (Edelstein effect) can occur in chiral crystals.
This is because an invariant coupling $G_{0}(\bm{T}\cdot\bm{M})$ exists in the free energy of chiral crystals with finite $G_{0}$.
Indeed, the Edelstein effect~\cite{Yoda2015, Yoda2018} and current-induced optical activity~\cite{1979JETPL, Shalygin2012} have been observed in elemental Te crystal~\cite{Furukawa2017, Furukawa2021}.
In addition, as the symmetry of $\bm{T}$ is common as a wave vector $\bm{k}$, the hedgehog spin texture observed around the H point of the Brillouin zone in Te~\cite{Hirayama2015, Sakano2020} is nothing but a consequence of the $\bm{k}$-space representation of the ET monopole, $\bm{k} \cdot \bm{\sigma}$~\cite{SH_MY_YY_HK_mul_2018}.

\begin{figure}[t!]
  \begin{center}
    \includegraphics[width=0.85 \hsize]{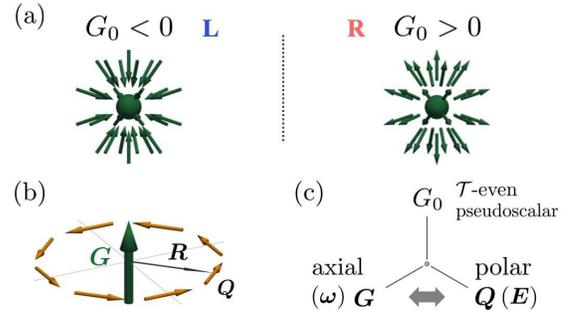}
    \vspace*{0mm}
    \caption{
      \label{fig:fig1}
      (a) ET monopole $G_{0}$ in terms of the flux structure of the ET dipoles $\bm{G}$, whose direction determines the handedness of the chirality.
      (b) Classical view of the ET dipole, which is a vortex-like alignment of the E dipoles, $\bm{Q}$.
      (c) The essential coupling existing in chiral crystals.
      The conjugate fields of $\bm{G}$ and $\bm{Q}$ are given in the parenthesis.
    }
  \end{center}
\end{figure}

An ET monopole $G_{0}$ can be decomposed in another way:
\begin{align}
  G_{0} \,\,\,\to\,\,\, \bm{Q} \cdot \bm{G},
  \label{eq:G0_QG}
\end{align}
where $\bm{Q}$ is an electric (E) dipole, $(\mathcal{P},\mathcal{T})=(-,+)$, such as a position vector $\bm{R}$, while $\bm{G}$ represents an ET dipole with $(\mathcal{P}, \mathcal{T}) = (+,+)$~\cite{SH_MY_YY_HK_mul_2018}.
Equation~(\ref{eq:G0_QG}) implies that $\bm{G}$-flux structure exists in chiral crystals as shown in Fig.\ref{fig:fig1}(a), and the direction of the fluxes characterizes the handedness.
Note that $\bm{G}$ vector solely plays a role of the order parameter of the recently discovered ferro-axial (rotational) order in RbFe(MoO$_{4}$)$_{2}$~\cite{Jin2020} and BaTiO$_{3}$~\cite{Hayashida2020}, and the anti-ferro-axial ordering in Ba(TiO)Cu$_{4}$(PO$_{4}$)$_{4}$~\cite{Hayashida2021}.
The rotational-ordered systems are closely related to chiral systems through Eq.~(\ref{eq:G0_QG}).

Since a classical representation of $\bm{G}$ is given by a vortex-like alignment of E dipoles $\bm{Q}$ as shown in Fig.~\ref{fig:fig1}(b), its conjugate field is a rotation of electric field, $\bm{\omega}_{\rm E}=\bm{\nabla}\times\bm{E}$, which is equivalent to time-dependent magnetic field through Maxwell's equation.
Moreover, from the symmetry property of $\bm{G}$, a lattice rotation $\bm{\omega}=\bm{\nabla} \times \bm{u}$ ($\bm{u}$ is a displacement vector of atoms) could be an alternative conjugate field to $\bm{G}$, provided a proper electron-lattice coupling.
From this relation, $\bm{G}$ is also significant in the transverse rotational phonon modes in both achiral~\cite{Zhang2015, Chen2018, Zhu2018} and chiral crystals~\cite{Kishine2020, Chen2021, ishito2021truly}.

With the above consideration, we find that chiral crystals substantially possess the third-order coupling in the free energy as shown in Fig.\ref{fig:fig1}(c):
\begin{align}
  F^{(3)} = g_{\perp} G_{0}^{(1)} (G_{x}Q_{x}+G_{y}Q_{y}) + g_{z} G_{0}^{(2)} G_{z} Q_{z},
  \label{eq:F3}
\end{align}
where the coupling constants satisfy $g_{z} = g_{\perp}$ in cubic crystals, otherwise $g_{z} \neq g_{\perp}$ ($z$ is along the screw axis).
Here, $G_{0}^{(1)}$ and $G_{0}^{(2)}$ can be independent ET monopoles in general.
This coupling gives rise to a conversion from a polar field such as the electric field or temperature gradient into an axial response such as a rotation of the lattice, and vice versa.
In other words, an electric-field induced rotation (EIR) and its inverse response, i.e., a rotation-field induced electric polarization (RIP), could appear in chiral crystals.
Note that one can apply a lattice rotation field by using transverse ultrasonic wave as it generates both the strain and rotation fields.
In this context, it is proposed that the temperature gradient gives rise to the lattice rotation in Te~\cite{Hamada2018}.

In what follows, we elucidate the microscopic origin of $G_{0}$ and related responses by taking Te as a specific example.
First, we construct the realistic tight-binding (TB) model of Te by using the results of the density-functional (DF) calculation.
Since the obtained TB model is expressed in terms of the symmetry-adopted electronic multipole basis, we can easily identify the microscopic origin of $G_{0}$, and evaluate the relevant couplings to it quantitatively.
We then propose a possible experimental approach to achieve absolute enantioselection, keeping the coupling, Eq.~(\ref{eq:F3}), in mind.

\begin{figure}[t!]
  \begin{center}
    \includegraphics[width=0.8 \hsize]{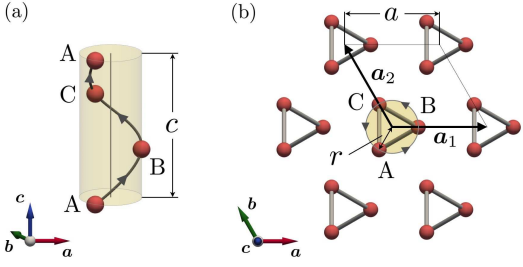}
    \vspace*{0mm}
    \caption{
    \label{fig:fig2}
    (a) Crystal structure of the right-handed Te.
    (b) A unit cell contains A, B, and C sublattices along a helical chain.
    }
  \end{center}
\end{figure}

%%%%%%%%%%%%%%%%%%%%%%%%%%%%%%%%%%%%%%%%%%%%%%%%%%%%%
{\it Tight-binding model for Te.---}
\label{sec:tb}
Let us first consider the specific TB model for Te.
As shown in Fig.~\ref{fig:fig2}(a), the bulk Te crystal consists of the threefold-symmetric helical chains, which contain A, B, and C sublattices in a unit cell as shown in Fig.~\ref{fig:fig2}(b).
The space group of the right- and left-handed Te are P3$_{1}$21 (\#152, $D_{3}^{4}$) and P3$_{2}$21 (\#154, $D_{3}^{6}$), respectively.
Hereafter, we focus on the right-handed Te.
The lattice constants are $a = 4.458$~\AA, $c = 5.925$~\AA, and relaxed value is $u = r/a = 0.274$ for the dimensionless helix parameter~\cite{Bouad_2003}, where $r$ denotes the radius of the helix.

Since the electronic states near the band edges (Fermi level) in Te mainly consist of three $p$ orbitals, $p_{x}$, $p_{y}$, and $p_{z}$~\cite{Cheng_2019}, we consider the spinful TB Hamiltonian in 18$\times$18 matrix.
The TB Hamiltonian is constructed by using the symmetry-adopted multipole basis $Z_{\alpha}$~\cite{SH_MY_YY_HK_mul_2018, HW_YY_Mul_2018, HK_RO_SH_comp_mul_2020} as $\mathcal{H} = \sum_{\alpha} z_{\alpha} Z_{\alpha}$, where $Z_{\alpha}$ are the independent multipole basis, which satisfy the normalization $\mathrm{Tr}(Z_{\alpha}Z_{\alpha}) = 1$, and belonging to the identity irreducible representation of $D_{3}^{4}$.
Each $Z_{\alpha}$ is expressed by the direct product of the cluster or bond basis~\cite{MTS_cmul_ahe_2017, MTS_fp_magmul_2018} and atomic basis as shown later~\cite{SH_bottom_up_2020, ROHK_2022}, which are generated automatically by the symbolic computation with symmetry operations~\cite{sm_comment}.
Then, the coefficients $z_{\alpha}$ ($\Delta_{i}^{Q}$, $\lambda_{i}^{Q}$, $t_{i}^{Q}$, $\alpha_{i}^{Q}$, ... etc.)
are determined so as to reproduce the band dispersions obtained by the DF calculation with the help of the machine-learning technique~\cite{Wang_2021,sm_comment}.
In this model construction, terms containing $G_{0}$ are particularly important.

The constructed TB Hamiltonian is expressed as
$  \mathcal{H}_{0}
 = \mathcal{H}_{\rm CEF} + \mathcal{H}_{\rm SOC} + \sum_{i = 1}^{8} \mathcal{H}_{t}^{(i)}$,
where $\mathcal{H}_{\rm CEF}$ and $\mathcal{H}_{\rm SOC}$ are the crystalline electric field (CEF) and spin-orbit coupling (SOC) within the unit cell, and $\mathcal{H}_{t}^{(i)}$ is the $i$-th neighbor hopping term.
By taking up to the 8th neighbors, we achieved the accuracy less than $10^{-4}$ of the mean squared error between the normalized energy eigenvalues of our TB model and DF result~\cite{sm_comment}.
The comparison of the energy dispersions is shown in Fig.~\ref{fig:fig4}(a).
We have also confirmed that the orbital and spin characters of the obtained electronic states roughly reproduce those obtained by DF calculation.
There are 255 independent parameters $z_{\alpha}$ in total, and 30 parameters are within the nearest-neighbor (NN) hopping; four CEF parameters, five SOC parameters, and twenty one NN intra-chain hopping parameters~\cite{sm_comment}.

\begin{figure}[t!]
  \begin{center}
    \includegraphics[width=1.0 \hsize]{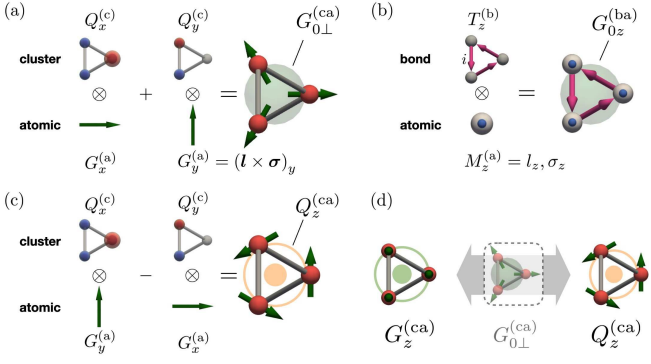}
    \vspace*{0mm}
    \caption{
      \label{fig:fig3}
      Schematic picture of multipole basis for Te.
      (a) local ET monopole, $G_{0\perp}^{\rm (ca)}$, (b) itinerant ET monopole, $G_{0z}^{\rm (ba)}$, and (c) local E dipole, $Q_{z}^{\rm (ca)}$.
      (d) The inter-parity coupling between the ET dipole $G_{z}^{\rm (ca)}$ and E dipole $Q_{z}^{\rm (ca)}$ via $G_{0\perp}^{\rm (ca)}$.
    }
  \end{center}
\end{figure}

\begin{figure*}[t!]
  \begin{center}
    \includegraphics[width=0.9 \hsize]{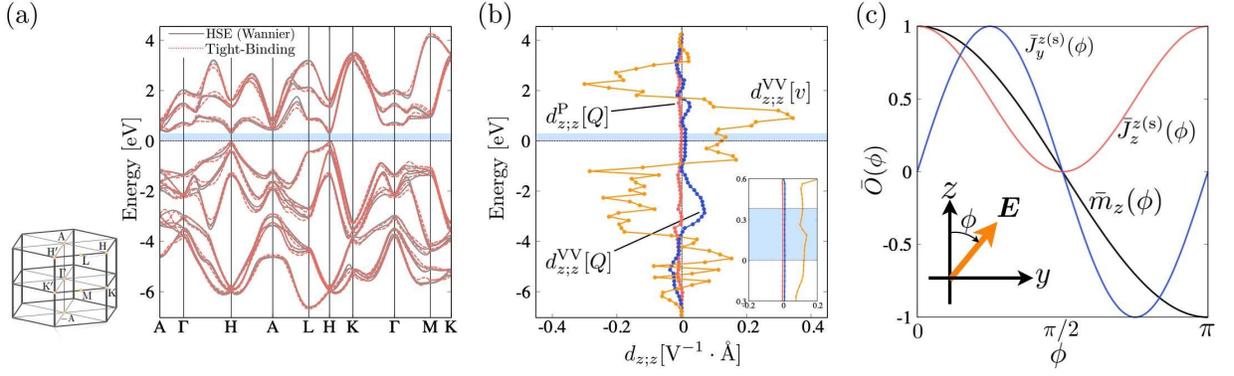}
    \vspace*{4mm}
    \caption{
    \label{fig:fig4}
    (a) The comparison of the band dispersion between our TB model and DF calculation.
    The Fermi-energy is taken as the origin, and the blue shaded area represents the insulating gap.
    (b) Chemical potential $\mu$ dependence of $d_{z;z}^{\rm P}$ and $d_{z;z}^{\rm VV}$ with Eqs.~(\ref{eq:dzz_P}), (\ref{eq:dzz_VV1}), and (\ref{eq:dzz_VV2}) at $T = 0.01$~eV, and $N=64^{3}$.
    The inset shows the enlarged plot near the Fermi level.
    (c) The electric-field angle dependence of the spin current in $yz$ plane $\bar{J}_{z}^{z {\rm (s)}}(\phi) \propto \cos^{2}(\phi)$ and $\bar{J}_{y}^{z {\rm (s)}}(\phi) \propto \sin(2\phi)$, and the magnetization due to the Edelstein effect, $\bar{m}_{z}(\phi) \propto \cos(\phi)$.
    $\bar{O}(\phi)$ denotes the normalized value of $O(\phi)$ by its absolute maximum value.
    }
  \end{center}
\end{figure*}

Among these multipole basis, the most dominant contributions containing the ET monopole are $G_{0\perp}^{\rm (ca)}$ in $\mathcal{H}_{\rm SOC}$ and $G_{0z}^{\rm (ba)}$, $G_{0\perp}^{\rm (ba)}$ in $\mathcal{H}_{t}^{(1)}$, which are given by
\begin{align}
   & G_{0\perp}^{\rm (ca)} = \frac{1}{\sqrt{2}} \left(Q_{x}^{\rm (c)} \otimes G_{x}^{\rm (a)} + Q_{y}^{\rm (c)} \otimes G_{y}^{\rm (a)}\right),
  \label{eq:G0p_ca}                                                                                                                             \\
   & G_{0z}^{\rm (ba)} = T_{z}^{\rm (b)} \otimes \sigma_{z}^{\rm (a)},
  \label{eq:G0z_ba} \\
   & G_{0\perp}^{\rm (ba)} = \frac{1}{\sqrt{2}} \left(T_{x}^{\rm (b)} \otimes \sigma_{x}^{\rm (a)} + T_{y}^{\rm (b)} \otimes \sigma_{y}^{\rm (a)}\right),
\end{align}
where the superscripts (c,b,a) represent cluster, bond, and atomic basis, respectively.
The weight $z_{\alpha}$ of $G_{0\perp}^{\rm (ca)}$ is $\lambda_{1}^{G} = 1.718$ eV which is the most dominant contribution among $\mathcal{H}_{\rm CEF}+\mathcal{H}_{\rm SOC}$, and that of $G_{0z}^{\rm (ba)}$ is $\alpha_{2}^{G} = 1.749$ eV which is the most dominant contribution among the ET monopoles in $\mathcal{H}_{t}^{(t)}$, while that of $G_{0\perp}^{\rm (ba)}$ is $\alpha_{3}^{G}=0.5854$ eV.
Note that $G_{0z}^{\rm (ca)}$ does not appear because the $z$ component of the E dipole $Q_{z}^{\rm (c)}$ identically vanishes in the present Hilbert space~\cite{sm_comment}.
We have confirmed that the magnitude of the parameters decreases for further neighbor hoppings~\cite{sm_comment}.
Here, $\bm{G}^{\rm (a)}=\bm{l}\times\bm{\sigma}$ is the atomic ET dipole ($\bm{l}$ and $\bm{\sigma}$ are the dimensionless orbital and half of spin angular momenta, respectively).
$Q_{\mu}^{\rm (c)}$ ($\mu = x, y$) and $T_{\mu}^{\rm (b)}$ ($\mu = x, y, z$) are the $\mu$ component of the cluster E dipole and bond MT dipole, which are defined in the ABC sublattice space as
\begin{align}
  Q_{x}^{\rm (c)} &=
  \frac{1}{\sqrt{6}}
  \begin{pmatrix}
    -1 & 0 & 0  \\
    0  & 2 & 0  \\
    0  & 0 & -1
  \end{pmatrix},\,\,\,
  Q_{y}^{\rm (c)} =
  \frac{1}{\sqrt{2}}
  \begin{pmatrix}
    -1 & 0 & 0 \\
    0  & 0 & 0 \\
    0  & 0 & 1
  \end{pmatrix}, \\
  Q_{0}^{\rm (c)} & =
  \frac{1}{\sqrt{3}}
  \begin{pmatrix}
    1 & 0 & 0 \\
    0 & 1 & 0 \\
    0 & 0 & 1
  \end{pmatrix},\,\,\,
  T_{z}^{\rm (b)} =
  \frac{1}{\sqrt{6}}
  \begin{pmatrix}
    0  & -i & i  \\
    i  & 0  & -i \\
    -i & i  & 0
  \end{pmatrix}, \\
  T_{x}^{\rm (b)} & =
  \frac{1}{2}
  \begin{pmatrix}
    0  & -i & 0  \\
    i  & 0  & i  \\
    0  & -i & 0
  \end{pmatrix},\,\,\,
  T_{y}^{\rm (b)} =
  \frac{1}{2\sqrt{3}}
  \begin{pmatrix}
    0  & -i & -2i \\
    i  & 0  & -i \\
    2i & i  & 0
  \end{pmatrix}.
\end{align}
Although there are also the spinless version of $G_{0z}^{\rm (ba)}$ and $G_{0\perp}^{\rm (ba)}$ which are given by replacing $\bm{\sigma}$ with $\bm{l}$ in Eq.~(\ref{eq:G0z_ba}), the weight of these multipole basis is much smaller than that for the spinful ones.
As shown in Fig.~\ref{fig:fig3}(a), $G_{0\perp}^{\rm (ca)}$ is the local ET monopole having the $\bm{G}^{\rm (a)}$-flux structure in the unit cell as Eq.~(\ref{eq:G0_QG}).
On the other hand, $G_{0z}^{\rm (ba)}$ in Fig.~\ref{fig:fig3}(b) is the itinerant ET monopole which is a kind of SOC with the spin-dependent imaginary hopping.
From Figs.~\ref{fig:fig3}(a), (b), the relation between the sign of $G_{0}$ and the handedness is apparent.
For example, in Fig.~\ref{fig:fig3}(b), $T_{z}^{\rm (b)}$ corresponds to the imaginary hopping directed to $+z$ direction.
If we consider the left-handed system, $T_{z}^{\rm (b)}$ directs to $-z$ direction, and the sign of $G_{0}$ is inverted.

These ET monopoles, $G_{0\perp}^{\rm (ca)}$ and $G_{0z}^{\rm (ba)}$, play roles of $G_{0}^{(1)}$ and $G_{0}^{(2)}$ in Eq.~(\ref{eq:F3}), respectively.
As discussed below, both of them give dominant contributions to the EIR and RIP.
In addition, the itinerant ET monopole $G_{0z}^{\rm (ba)}$ is also the main origin of the Edelstein effect observed in Te~\cite{Furukawa2017, Furukawa2021}.
Moreover, the Fourier transform of $G_{0z}^{\rm (ba)}$ together with $G_{0\perp}^{\rm (ba)}$ give rise to the hedgehog spin texture around the H point in the momentum space~\cite{Hirayama2015, Sakano2020, sm_comment}.

%%%%%%%%%%%%%%%%%%%%%%%%%%%%%%%%%%%%%%%%%%%%%%%%%%%%%
{\it Electric-field induced rotation.---}
\label{sec:eir}
As already discussed phenomenologically, a conversion between polar and axial degrees of freedom is expected to occur in chiral materials via Eq.~(\ref{eq:F3}).
To demonstrate it microscopically, we investigate the electric-field induced ET dipole response based on the present TB model.
Before showing the results, we define explicitly the local E- and ET-dipole operators that describe the input and output of the response.
The E dipole $Q_{z}^{\rm (ca)}$ is given by the similar expression of Eq.~(\ref{eq:G0p_ca}) with a minus sign for the second term, which is schematically shown in Fig.~\ref{fig:fig3}(c), i.e., the vortex-like alignment of $G_{x}^{\rm (a)}$ and $G_{y}^{\rm (a)}$ (cf. the roles of $\bm{G}$ and $\bm{Q}$ are reverted in Fig.~\ref{fig:fig1}(b)).
The output rotation is also described by $G_{z}^{\rm (ca)} = Q_{0}^{\rm (c)} \otimes G_{z}^{\rm (a)}$.

Using the Kubo formula, the response function in $G_{z}^{\rm (ca)} = d_{z;z} E_{z}$ is expressed as ($c$ is the lattice constant)
\begin{align}
   d_{z;z} &= d_{z;z}^{\rm P}[Q] + d_{z;z}^{\rm VV}[Q] + d_{z;z}^{\rm VV}[v],
   \label{eq:dzz}\\
   &d_{z;z}^{\rm P}[Q] = -\frac{ec}{N} \sum_{\bm{k} nm}^{\epsilon_{n\bm{k}} = \epsilon_{m\bm{k}}} \frac{\partial f_{n\bm{k}}}{\partial \epsilon_{n\bm{k}}} G_{z\bm{k}}^{nm} Q_{z\bm{k}}^{mn},
   \label{eq:dzz_P}\\
    &d_{z;z}^{\rm VV}[Q]
     = \frac{ec}{N} \sum_{\bm{k} nm}^{\epsilon_{n\bm{k}} \neq \epsilon_{m\bm{k}}}  \frac{f_{n\bm{k}} - f_{m\bm{k}}}{\epsilon_{n\bm{k}} - \epsilon_{m\bm{k}}} G_{z\bm{k}}^{nm} Q_{z\bm{k}}^{mn},
   \label{eq:dzz_VV1}\\
    &d_{z;z}^{\rm VV}[v]
    = -\frac{e \hbar}{iN} \sum_{\bm{k} nm}^{\epsilon_{n\bm{k}} \neq \epsilon_{m\bm{k}}} \frac{f_{n \bm{k}} - f_{m \bm{k}}}{(\epsilon_{n\bm{k}} - \epsilon_{m\bm{k}})^{2}} G_{z\bm{k}}^{n m} v_{z\bm{k}}^{m n}.
   \label{eq:dzz_VV2}
\end{align}
Here, the matrix element of an operator $\hat{O}$ is $O_{i\bm{k}}^{nm}=\braket{\psi_{n\bm{k}}|\hat{O}_{i}|\psi_{m\bm{k}}}$, $f_{n\bm{k}} = f(\epsilon_{n\bm{k}})$ is the Fermi distribution function, $e$ $(>0)$ is the elementary charge, and $N$ is the number of lattice sites.
The responses $d_{z;z}^{\rm P}$ and $d_{z;z}^{\rm VV}$ represent the intra-band Pauli contribution proportional to the density of states (DOS), and inter-band van Vleck contributions, respectively.
$[Q]$ and $[v]$ indicate the contributions arising from the local E dipole and itinerant hopping process via the velocity operator, $\bm{v}_{\bm{k}}=\partial\mathcal{H}_{0}/\partial\hbar\bm{k}$, respectively.
Note that $d_{z;z}^{\rm P}[v]$ vanishes identically by the symmetry.
We have used $N=64^{3}$ and the temperature $T=0.01$ eV in the following results.

Figure \ref{fig:fig4}(b) shows the chemical potential $\mu$ dependence of the responses.
The inter-band contribution from the itinerant hopping process, $d_{z;z}^{\rm VV}[v]$, is always dominant irrespective of $\mu$, and the EIR occurs even in the insulator.
Analyzing the essential parameters to exhibit the finite response $d_{z;z}$~\cite{ROHK_2022}, it turns out that the lowest-order of the response is proportional to the highest-weight term $G_{0z}^{\rm (ba)}$ of $\mathcal{H}_{0}$ with the coefficient $\alpha_{2}^{G}$, which is consistent with the fact that $d_{z;z}^{\rm VV}[v]$ is dominant in numerical result.
Thus, the itinerant ET monopole $G_{0z}^{\rm (ba)}$ is the key component of the EIR response in Te.
Note that the inverse RIP process is also expected to occur in both metals and insulators, as their response functions are common with Eqs.~(\ref{eq:dzz_P}) and (\ref{eq:dzz_VV1}).

Although we have concentrated on the electronic responses in the above, the actual lattice rotation should occur via the electron-lattice coupling.
When we restrict our discussion to the pure lattice rotation with the angle $\omega_{z}$ with respect to $z$ screw axis, the electron-lattice coupling can be evaluated by rotating inversely the electronic system by the angle $-\omega_{z}$~\cite{Goto_1986, Kuromaru_2000, Kurihara_2017}. Namely,
$  \mathcal{H}^{(z)}_{\text{el-rot}}
   = e^{-ij_{z}^{\rm (ca)}\omega_{z}} \mathcal{H}_{0} e^{ij_{z}^{\rm (ca)}\omega_{z}}  - \mathcal{H}_{0}
  = i[\mathcal{H}_{0},j_{z}^{\rm (ca)}] \omega_{z} + \cdots$,
where $j_{z}^{\rm (ca)} = Q_{0}^{\rm (c)} \otimes (l_{z} + \sigma_{z}/2)$ is the total angular momentum.
We find the most important contribution from $\lambda_{1}^{G} G_{0\perp}^{\rm (ca)}$ term in $\mathcal{H}_{0}$ as
\begin{align}
  \mathcal{H}^{(z)}_{\text{el-rot}} \sim \lambda_{1}^{G} Q_{z}^{\rm (ca)} \omega_{z},
  \label{eq:H_ER_z}
\end{align}
with $\lambda_{1}^{G}=1.718$ eV.
Similarly, the perpendicular components are obtained, and they are a factor $1/\sqrt{2}$ smaller than $\mathcal{H}^{(z)}_{\text{el-rot}}$.
This term causes the electric polarization directly by applying a lattice rotation field with the use of transverse ultrasound wave for instance.

The induced electronic ET dipole can also be observed by the spin-current measurement.
When the induced ET dipole by the electric field $E_{z}$ is present, two types of nonlinear spin currents are expected: $J_{z}^{z {\rm (s)}} = \sigma_{z;zz}^{z{\rm (s)}} E_{z}^{2}$ and $J_{y}^{z {\rm (s)}} = \sigma_{y;yz}^{z{\rm (s)}} E_{y} E_{z}$, where $J_{\mu}^{\nu {\rm (s)}} \equiv (J_{\mu}\sigma_{\nu}+\sigma_{\nu}J_{\mu})/2$ is the spin current.
As shown in Fig.~\ref{fig:fig4}(c), the electric-field angle $\phi$ dependences of $J_{\mu}^{\nu {\rm (s)}}$ in $yz$ plane are given by $J_{z}^{z {\rm (s)}}(\phi) \propto \cos^{2}(\phi)$ and $J_{y}^{z {\rm (s)}}(\phi) \propto \sin(2\phi)$, respectively.
Note that they are in marked contrast to that of the magnetization due to the Edelstein effect, $m_{z}(\phi) = \alpha_{z;z}^{\rm (J)} E_{z} \propto \cos(\phi)$.
Thus, it is verified whether the ET dipole is induced by the electric field or not, by the $\phi$ dependence of $J_{z}^{z {\rm (s)}}(\phi)$ and $J_{y}^{z {\rm (s)}}(\phi)$.

\begin{figure}[b!]
  \begin{center}
    \includegraphics[width=0.9 \hsize]{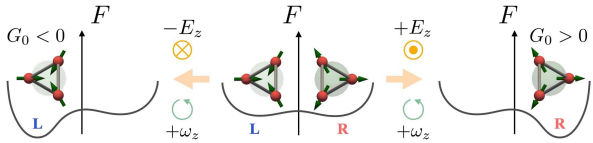}
    \vspace*{0mm}
    \caption{
      \label{fig:fig5}
      Free-energy differences for absolute enantioselection under simultaneous application of rotation (${\omega}_{z}$) and electric $(E_{z}$) fields. The sign of $\omega_{z}E_{z}$ controls the preferred handedness during crystallization process.
    }
  \end{center}
\end{figure}

%%%%%%%%%%%%%%%%%%%%%%%%%%%%%%%%%%%%%%%%%%%%%%%%%%%%%
{\it Absolute enantioselection by rotation and electric fields.---}
\label{sec:cc}
Finally, we propose a possible experimental approach to achieve absolute enantioselection in chiral materials.
As was shown phenomenologically and microscopically, there is a proper coupling among $G_{0}$, $\bm{G}$, and $\bm{Q}$ in chiral materials.
The conjugate field of $\bm{Q}$ is an electric field $\bm{E}$ and that of $\bm{G}$ is a rotation field $\bm{\omega}$ such as a rotation of electric field or equivalently time-dependent magnetic field.
Thus, the conjugate field of the chirality, that is $G_{0}$, is a product of polar and axial vector fields, $E_{\mu}$ and $\omega_{\mu}$.
In the electromagnetism, it is known as the quantity called zilch, which describes optical chirality\cite{proskurin_2017}.
Therefore, as shown in Fig.~\ref{fig:fig5} in the case of $\mu = z$, absolute enantioselection can be achieved by simultaneous application of electric $(E_{\mu}$) and rotation ($\omega_{\mu}$) fields for instance.
The sign of $\omega_{\mu}E_{\mu}$ controls the preferred handedness during crystallization process, as shown in the left-most and right-most panels in Fig.~\ref{fig:fig5}.
It should be emphasized that a time-dependent magnetic field is favorable with the constant time derivative or net accumulation with definite sign.
In addition to this, both electric and magnetic fields must be parallel with each other.
Moreover, as was discussed in Eq.~(\ref{eq:G0_TM}), $G_{0}$ could also couple with $M_{\mu}T_{\mu}$.
Thus, the combined static magnetic field and time-dependent electric field (i.e., $\bm{\nabla}\times\bm{B}$, which is a conjugate field of $T_{\mu}$), or the combined static magnetic field and electric current, can be used to achieve absolute enantioselection as well.
This generic approach is applicable to any chiral material.

%%%%%%%%%%%%%%%%%%%%%%%%%%%%%%%%%%%%%%%%%%%%%%%%%%%%%
{\it Conclusion.---}
\label{sec:conclusion}
We have unveiled the microscopic origin of chirality and possible electric-field induced rotation and its inverse response, rotation-field induced electric polarization.
First, we have demonstrated that the chirality corresponds to the electric toroidal monopole given by Eq.~(\ref{eq:G0_QG}), in terms of symmetry.
We have also found the essential coupling to it as shown in Eq.~(\ref{eq:F3}), which is the key element for both the electric-field induced rotation and its inverse response.
Then, using the realistic tight-binding model for elemental Te crystal, we have elucidated that the inter-band process, driven by the itinerant electric toroidal monopole shown in Fig.~\ref{fig:fig3}(b), is the crucial factor in these response functions.
These responses occur even in the insulators, which is in marked contrast to the Edelstein effect observed in Te.
Lastly, we have proposed a generic experimental approach to realize the absolute enantioselection in chiral materials by the conjugate field of the chirality, such as simultaneously applied electric and rotation fields, or magnetic field and electric current, and so on.

%%%%%%%%%%%%%%%%%%%%%%%%%%%%%%%%%%%%%%%%%%%%%%%%%%%%%
% acknowledgments
The authors would like to thank Yuki Yanagi for providing us the data of the DF calculation.
We also thank Michito Suzuki, Hiroaki Ikeda, Yoshihiko Togawa, Junichiro Kishine, Koichi Izawa, Hiroshi Amitsuka, Tatsuya Yanagisawa, Hiroyuki Hidaka, and Satoru Hayami for fruitful discussions.
This work was supported by JSPS KAKENHI Grants Numbers JP19K03752, JPJP20J21838.
A part of numerical and symbolic calculations was performed in the supercomputing systems in the MAterial science Supercomputing system for Advanced MUlti-scale simulations towards NExt-generation-Institute for Materials Research (MASAMUNE-IMR) of the Center for Computational Materials Science, Institute for Materials Research, Tohoku University.

%%%%%%%%%%%%%%%%%%%%%%%%%%%%%%%%%%%%%%%%%%%%%%%%%%%%%
%\bibliography{apssamp}% Produces the bibliography via BibTeX.
%\bibliographystyle{junsrt}
\bibliographystyle{apsrev4-2}
% \bibliography{te_v13}

\begin{thebibliography}{36}%
  \makeatletter
  \providecommand \@ifxundefined [1]{%
   \@ifx{#1\undefined}
  }%
  \providecommand \@ifnum [1]{%
   \ifnum #1\expandafter \@firstoftwo
   \else \expandafter \@secondoftwo
   \fi
  }%
  \providecommand \@ifx [1]{%
   \ifx #1\expandafter \@firstoftwo
   \else \expandafter \@secondoftwo
   \fi
  }%
  \providecommand \natexlab [1]{#1}%
  \providecommand \enquote  [1]{``#1''}%
  \providecommand \bibnamefont  [1]{#1}%
  \providecommand \bibfnamefont [1]{#1}%
  \providecommand \citenamefont [1]{#1}%
  \providecommand \href@noop [0]{\@secondoftwo}%
  \providecommand \href [0]{\begingroup \@sanitize@url \@href}%
  \providecommand \@href[1]{\@@startlink{#1}\@@href}%
  \providecommand \@@href[1]{\endgroup#1\@@endlink}%
  \providecommand \@sanitize@url [0]{\catcode `\\12\catcode `\$12\catcode
    `\&12\catcode `\#12\catcode `\^12\catcode `\_12\catcode `\%12\relax}%
  \providecommand \@@startlink[1]{}%
  \providecommand \@@endlink[0]{}%
  \providecommand \url  [0]{\begingroup\@sanitize@url \@url }%
  \providecommand \@url [1]{\endgroup\@href {#1}{\urlprefix }}%
  \providecommand \urlprefix  [0]{URL }%
  \providecommand \Eprint [0]{\href }%
  \providecommand \doibase [0]{https://doi.org/}%
  \providecommand \selectlanguage [0]{\@gobble}%
  \providecommand \bibinfo  [0]{\@secondoftwo}%
  \providecommand \bibfield  [0]{\@secondoftwo}%
  \providecommand \translation [1]{[#1]}%
  \providecommand \BibitemOpen [0]{}%
  \providecommand \bibitemStop [0]{}%
  \providecommand \bibitemNoStop [0]{.\EOS\space}%
  \providecommand \EOS [0]{\spacefactor3000\relax}%
  \providecommand \BibitemShut  [1]{\csname bibitem#1\endcsname}%
  \let\auto@bib@innerbib\@empty
  %</preamble>
  \bibitem [{\citenamefont {Barron}(2004)}]{barron_2004}%
    \BibitemOpen
    \bibfield  {author} {\bibinfo {author} {\bibfnamefont {L.~D.}\ \bibnamefont
    {Barron}},\ }\href {https://doi.org/10.1017/CBO9780511535468} {\emph
    {\bibinfo {title} {Molecular Light Scattering and Optical Activity}}},\
    \bibinfo {edition} {2nd}\ ed.\ (\bibinfo  {publisher} {Cambridge University
    Press},\ \bibinfo {year} {2004})\BibitemShut {NoStop}%
  \bibitem [{\citenamefont {Barron}(2013)}]{barron_2013}%
    \BibitemOpen
    \bibfield  {author} {\bibinfo {author} {\bibfnamefont {L.~D.}\ \bibnamefont
    {Barron}},\ }\href@noop {} {\bibfield  {journal} {\bibinfo  {journal} {Rend.
    Fis. Acc. Lincei}\ }\textbf {\bibinfo {volume} {24}},\ \bibinfo {pages} {179}
    (\bibinfo {year} {2013})}\BibitemShut {NoStop}%
  \bibitem [{\citenamefont {Hayami}\ and\ \citenamefont{Kusunose}(2018)}]{SH_HK_2018}%
    \BibitemOpen
    \bibfield  {author} {\bibinfo {author} {\bibfnamefont {S.}~\bibnamefont
    {Hayami}}\ and\ \bibinfo {author} {\bibfnamefont {H.}~\bibnamefont
    {Kusunose}},\ }\href {https://doi.org/10.7566/JPSJ.87.033709} {\bibfield
    {journal} {\bibinfo  {journal} {J. Phys. Soc. Jpn.}\ }\textbf {\bibinfo
    {volume} {87}},\ \bibinfo {pages} {033709} (\bibinfo {year}
    {2018})}\BibitemShut {NoStop}%
  \bibitem [{\citenamefont {Hayami}\ \emph {et~al.}(2018)\citenamefont {Hayami},
    \citenamefont {Yatsushiro}, \citenamefont {Yanagi},\ and\ \citenamefont
    {Kusunose}}]{SH_MY_YY_HK_mul_2018}%
    \BibitemOpen
    \bibfield  {author} {\bibinfo {author} {\bibfnamefont {S.}~\bibnamefont
    {Hayami}}, \bibinfo {author} {\bibfnamefont {M.}~\bibnamefont {Yatsushiro}},
    \bibinfo {author} {\bibfnamefont {Y.}~\bibnamefont {Yanagi}},\ and\ \bibinfo
    {author} {\bibfnamefont {H.}~\bibnamefont {Kusunose}},\ }\href
    {https://doi.org/10.1103/PhysRevB.98.165110} {\bibfield  {journal} {\bibinfo
    {journal} {Phys. Rev. B}\ }\textbf {\bibinfo {volume} {98}},\ \bibinfo
    {pages} {165110} (\bibinfo {year} {2018})}\BibitemShut {NoStop}%
  \bibitem [{\citenamefont {Watanabe}\ and\ \citenamefont
    {Yanase}(2018)}]{HW_YY_Mul_2018}%
    \BibitemOpen
    \bibfield  {author} {\bibinfo {author} {\bibfnamefont {H.}~\bibnamefont
    {Watanabe}}\ and\ \bibinfo {author} {\bibfnamefont {Y.}~\bibnamefont
    {Yanase}},\ }\href {https://doi.org/10.1103/PhysRevB.98.245129} {\bibfield
    {journal} {\bibinfo  {journal} {Phys. Rev. B}\ }\textbf {\bibinfo {volume}
    {98}},\ \bibinfo {pages} {245129} (\bibinfo {year} {2018})}\BibitemShut
    {NoStop}%
  \bibitem [{\citenamefont {Kusunose}\ \emph {et~al.}(2020)\citenamefont
    {Kusunose}, \citenamefont {Oiwa},\ and\ \citenamefont
    {Hayami}}]{HK_RO_SH_comp_mul_2020}%
    \BibitemOpen
    \bibfield  {author} {\bibinfo {author} {\bibfnamefont {H.}~\bibnamefont
    {Kusunose}}, \bibinfo {author} {\bibfnamefont {R.}~\bibnamefont {Oiwa}},\
    and\ \bibinfo {author} {\bibfnamefont {S.}~\bibnamefont {Hayami}},\ }\href
    {https://doi.org/10.7566/JPSJ.89.104704} {\bibfield  {journal} {\bibinfo
    {journal} {J. Phys. Soc. Jpn.}\ }\textbf {\bibinfo {volume} {89}},\ \bibinfo
    {pages} {104704} (\bibinfo {year} {2020})}\BibitemShut {NoStop}%
  \bibitem [{\citenamefont {Yoda}\ \emph {et~al.}(2015)\citenamefont {Yoda},
    \citenamefont {Yokoyama},\ and\ \citenamefont {Murakami}}]{Yoda2015}%
    \BibitemOpen
    \bibfield  {author} {\bibinfo {author} {\bibfnamefont {T.}~\bibnamefont
    {Yoda}}, \bibinfo {author} {\bibfnamefont {T.}~\bibnamefont {Yokoyama}},\
    and\ \bibinfo {author} {\bibfnamefont {S.}~\bibnamefont {Murakami}},\ }\href
    {https://doi.org/10.1038/srep12024} {\bibfield  {journal} {\bibinfo
    {journal} {Sci. Rep.}\ }\textbf {\bibinfo {volume} {5}},\ \bibinfo {pages}
    {12024} (\bibinfo {year} {2015})}\BibitemShut {NoStop}%
  \bibitem [{\citenamefont {Yoda}\ \emph {et~al.}(2018)\citenamefont {Yoda},
    \citenamefont {Yokoyama},\ and\ \citenamefont {Murakami}}]{Yoda2018}%
    \BibitemOpen
    \bibfield  {author} {\bibinfo {author} {\bibfnamefont {T.}~\bibnamefont
    {Yoda}}, \bibinfo {author} {\bibfnamefont {T.}~\bibnamefont {Yokoyama}},\
    and\ \bibinfo {author} {\bibfnamefont {S.}~\bibnamefont {Murakami}},\ }\href
    {https://doi.org/10.1021/acs.nanolett.7b04300} {\bibfield  {journal}
    {\bibinfo  {journal} {Nano Lett.}\ }\textbf {\bibinfo {volume} {18}},\
    \bibinfo {pages} {916} (\bibinfo {year} {2018})}\BibitemShut {NoStop}%
  \bibitem [{\citenamefont {{Vorob'ev}}\ \emph {et~al.}(1979)\citenamefont
    {{Vorob'ev}}, \citenamefont {{Ivchenko}}, \citenamefont {{Pikus}},
    \citenamefont {{Farbshte{\v{i}}n}}, \citenamefont {{Shalygin}},\ and\
    \citenamefont {{Shturbin}}}]{1979JETPL}%
    \BibitemOpen
    \bibfield  {author} {\bibinfo {author} {\bibfnamefont {L.~E.}\ \bibnamefont
    {{Vorob'ev}}}, \bibinfo {author} {\bibfnamefont {E.~L.}\ \bibnamefont
    {{Ivchenko}}}, \bibinfo {author} {\bibfnamefont {G.~E.}\ \bibnamefont
    {{Pikus}}}, \bibinfo {author} {\bibfnamefont {I.~I.}\ \bibnamefont
    {{Farbshte{\v{i}}n}}}, \bibinfo {author} {\bibfnamefont {V.~A.}\ \bibnamefont
    {{Shalygin}}},\ and\ \bibinfo {author} {\bibfnamefont {A.~V.}\ \bibnamefont
    {{Shturbin}}},\ }\href@noop {} {\bibfield  {journal} {\bibinfo  {journal}
    {JETP Lett.}\ }\textbf {\bibinfo {volume} {29}},\ \bibinfo {pages} {441}
    (\bibinfo {year} {1979})}\BibitemShut {NoStop}%
  \bibitem [{\citenamefont {Shalygin}\ \emph {et~al.}(2012)\citenamefont
    {Shalygin}, \citenamefont {Sofronov}, \citenamefont {Vorob'ev},\ and\
    \citenamefont {Farbshtein}}]{Shalygin2012}%
    \BibitemOpen
    \bibfield  {author} {\bibinfo {author} {\bibfnamefont {V.~A.}\ \bibnamefont
    {Shalygin}}, \bibinfo {author} {\bibfnamefont {A.~N.}\ \bibnamefont
    {Sofronov}}, \bibinfo {author} {\bibfnamefont {L.~E.}\ \bibnamefont
    {Vorob'ev}},\ and\ \bibinfo {author} {\bibfnamefont {I.~I.}\ \bibnamefont
    {Farbshtein}},\ }\href {https://doi.org/10.1134/S1063783412120281} {\bibfield
     {journal} {\bibinfo  {journal} {Phys. Solid State}\ }\textbf {\bibinfo
    {volume} {54}},\ \bibinfo {pages} {2362} (\bibinfo {year}
    {2012})}\BibitemShut {NoStop}%
  \bibitem [{\citenamefont {Furukawa}\ \emph {et~al.}(2017)\citenamefont
    {Furukawa}, \citenamefont {Shimokawa}, \citenamefont {Kobayashi},\ and\
    \citenamefont {Itou}}]{Furukawa2017}%
    \BibitemOpen
    \bibfield  {author} {\bibinfo {author} {\bibfnamefont {T.}~\bibnamefont
    {Furukawa}}, \bibinfo {author} {\bibfnamefont {Y.}~\bibnamefont {Shimokawa}},
    \bibinfo {author} {\bibfnamefont {K.}~\bibnamefont {Kobayashi}},\ and\
    \bibinfo {author} {\bibfnamefont {T.}~\bibnamefont {Itou}},\ }\href
    {https://doi.org/10.1038/s41467-017-01093-3} {\bibfield  {journal} {\bibinfo
    {journal} {Nat. Commun.}\ }\textbf {\bibinfo {volume} {8}},\ \bibinfo {pages}
    {954} (\bibinfo {year} {2017})}\BibitemShut {NoStop}%
  \bibitem [{\citenamefont {Furukawa}\ \emph {et~al.}(2021)\citenamefont
    {Furukawa}, \citenamefont {Watanabe}, \citenamefont {Ogasawara},
    \citenamefont {Kobayashi},\ and\ \citenamefont {Itou}}]{Furukawa2021}%
    \BibitemOpen
    \bibfield  {author} {\bibinfo {author} {\bibfnamefont {T.}~\bibnamefont
    {Furukawa}}, \bibinfo {author} {\bibfnamefont {Y.}~\bibnamefont {Watanabe}},
    \bibinfo {author} {\bibfnamefont {N.}~\bibnamefont {Ogasawara}}, \bibinfo
    {author} {\bibfnamefont {K.}~\bibnamefont {Kobayashi}},\ and\ \bibinfo
    {author} {\bibfnamefont {T.}~\bibnamefont {Itou}},\ }\href
    {https://doi.org/10.1103/PhysRevResearch.3.023111} {\bibfield  {journal}
    {\bibinfo  {journal} {Phys. Rev. Research}\ }\textbf {\bibinfo {volume}
    {3}},\ \bibinfo {pages} {023111} (\bibinfo {year} {2021})}\BibitemShut
    {NoStop}%
  \bibitem [{\citenamefont {Hirayama}\ \emph {et~al.}(2015)\citenamefont
    {Hirayama}, \citenamefont {Okugawa}, \citenamefont {Ishibashi}, \citenamefont
    {Murakami},\ and\ \citenamefont {Miyake}}]{Hirayama2015}%
    \BibitemOpen
    \bibfield  {author} {\bibinfo {author} {\bibfnamefont {M.}~\bibnamefont
    {Hirayama}}, \bibinfo {author} {\bibfnamefont {R.}~\bibnamefont {Okugawa}},
    \bibinfo {author} {\bibfnamefont {S.}~\bibnamefont {Ishibashi}}, \bibinfo
    {author} {\bibfnamefont {S.}~\bibnamefont {Murakami}},\ and\ \bibinfo
    {author} {\bibfnamefont {T.}~\bibnamefont {Miyake}},\ }\href
    {https://doi.org/10.1103/PhysRevLett.114.206401} {\bibfield  {journal}
    {\bibinfo  {journal} {Phys. Rev. Lett.}\ }\textbf {\bibinfo {volume} {114}},\
    \bibinfo {pages} {206401} (\bibinfo {year} {2015})}\BibitemShut {NoStop}%
  \bibitem [{\citenamefont {Sakano}\ \emph {et~al.}(2020)\citenamefont {Sakano},
    \citenamefont {Hirayama}, \citenamefont {Takahashi}, \citenamefont {Akebi},
    \citenamefont {Nakayama}, \citenamefont {Kuroda}, \citenamefont {Taguchi},
    \citenamefont {Yoshikawa}, \citenamefont {Miyamoto}, \citenamefont {Okuda},
    \citenamefont {Ono}, \citenamefont {Kumigashira}, \citenamefont {Ideue},
    \citenamefont {Iwasa}, \citenamefont {Mitsuishi}, \citenamefont {Ishizaka},
    \citenamefont {Shin}, \citenamefont {Miyake}, \citenamefont {Murakami},
    \citenamefont {Sasagawa},\ and\ \citenamefont {Kondo}}]{Sakano2020}%
    \BibitemOpen
    \bibfield  {author} {\bibinfo {author} {\bibfnamefont {M.}~\bibnamefont
    {Sakano}}, \bibinfo {author} {\bibfnamefont {M.}~\bibnamefont {Hirayama}},
    \bibinfo {author} {\bibfnamefont {T.}~\bibnamefont {Takahashi}}, \bibinfo
    {author} {\bibfnamefont {S.}~\bibnamefont {Akebi}}, \bibinfo {author}
    {\bibfnamefont {M.}~\bibnamefont {Nakayama}}, \bibinfo {author}
    {\bibfnamefont {K.}~\bibnamefont {Kuroda}}, \bibinfo {author} {\bibfnamefont
    {K.}~\bibnamefont {Taguchi}}, \bibinfo {author} {\bibfnamefont
    {T.}~\bibnamefont {Yoshikawa}}, \bibinfo {author} {\bibfnamefont
    {K.}~\bibnamefont {Miyamoto}}, \bibinfo {author} {\bibfnamefont
    {T.}~\bibnamefont {Okuda}}, \bibinfo {author} {\bibfnamefont
    {K.}~\bibnamefont {Ono}}, \bibinfo {author} {\bibfnamefont {H.}~\bibnamefont
    {Kumigashira}}, \bibinfo {author} {\bibfnamefont {T.}~\bibnamefont {Ideue}},
    \bibinfo {author} {\bibfnamefont {Y.}~\bibnamefont {Iwasa}}, \bibinfo
    {author} {\bibfnamefont {N.}~\bibnamefont {Mitsuishi}}, \bibinfo {author}
    {\bibfnamefont {K.}~\bibnamefont {Ishizaka}}, \bibinfo {author}
    {\bibfnamefont {S.}~\bibnamefont {Shin}}, \bibinfo {author} {\bibfnamefont
    {T.}~\bibnamefont {Miyake}}, \bibinfo {author} {\bibfnamefont
    {S.}~\bibnamefont {Murakami}}, \bibinfo {author} {\bibfnamefont
    {T.}~\bibnamefont {Sasagawa}},\ and\ \bibinfo {author} {\bibfnamefont
    {T.}~\bibnamefont {Kondo}},\ }\href
    {https://doi.org/10.1103/PhysRevLett.124.136404} {\bibfield  {journal}
    {\bibinfo  {journal} {Phys. Rev. Lett.}\ }\textbf {\bibinfo {volume} {124}},\
    \bibinfo {pages} {136404} (\bibinfo {year} {2020})}\BibitemShut {NoStop}%
  \bibitem [{\citenamefont {Jin}\ \emph {et~al.}(2020)\citenamefont {Jin},
    \citenamefont {Drueke}, \citenamefont {Li}, \citenamefont {Admasu},
    \citenamefont {Owen}, \citenamefont {Day}, \citenamefont {Sun}, \citenamefont
    {Cheong},\ and\ \citenamefont {Zhao}}]{Jin2020}%
    \BibitemOpen
    \bibfield  {author} {\bibinfo {author} {\bibfnamefont {W.}~\bibnamefont
    {Jin}}, \bibinfo {author} {\bibfnamefont {E.}~\bibnamefont {Drueke}},
    \bibinfo {author} {\bibfnamefont {S.}~\bibnamefont {Li}}, \bibinfo {author}
    {\bibfnamefont {A.}~\bibnamefont {Admasu}}, \bibinfo {author} {\bibfnamefont
    {R.}~\bibnamefont {Owen}}, \bibinfo {author} {\bibfnamefont {M.}~\bibnamefont
    {Day}}, \bibinfo {author} {\bibfnamefont {K.}~\bibnamefont {Sun}}, \bibinfo
    {author} {\bibfnamefont {S.-W.}\ \bibnamefont {Cheong}},\ and\ \bibinfo
    {author} {\bibfnamefont {L.}~\bibnamefont {Zhao}},\ }\href
    {https://doi.org/10.1038/s41567-019-0695-1} {\bibfield  {journal} {\bibinfo
    {journal} {Nat. Phys.}\ }\textbf {\bibinfo {volume} {16}},\ \bibinfo {pages}
    {42} (\bibinfo {year} {2020})}\BibitemShut {NoStop}%
  \bibitem [{\citenamefont {Hayashida}\ \emph {et~al.}(2020)\citenamefont
    {Hayashida}, \citenamefont {Uemura}, \citenamefont {Kimura}, \citenamefont
    {Matsuoka}, \citenamefont {Morikawa}, \citenamefont {Hirose}, \citenamefont
    {Tsuda}, \citenamefont {Hasegawa},\ and\ \citenamefont
    {Kimura}}]{Hayashida2020}%
    \BibitemOpen
    \bibfield  {author} {\bibinfo {author} {\bibfnamefont {T.}~\bibnamefont
    {Hayashida}}, \bibinfo {author} {\bibfnamefont {Y.}~\bibnamefont {Uemura}},
    \bibinfo {author} {\bibfnamefont {K.}~\bibnamefont {Kimura}}, \bibinfo
    {author} {\bibfnamefont {S.}~\bibnamefont {Matsuoka}}, \bibinfo {author}
    {\bibfnamefont {D.}~\bibnamefont {Morikawa}}, \bibinfo {author}
    {\bibfnamefont {S.}~\bibnamefont {Hirose}}, \bibinfo {author} {\bibfnamefont
    {K.}~\bibnamefont {Tsuda}}, \bibinfo {author} {\bibfnamefont
    {T.}~\bibnamefont {Hasegawa}},\ and\ \bibinfo {author} {\bibfnamefont
    {T.}~\bibnamefont {Kimura}},\ }\href
    {https://doi.org/10.1038/s41467-020-18408-6} {\bibfield  {journal} {\bibinfo
    {journal} {Nat. Commun.}\ }\textbf {\bibinfo {volume} {11}},\ \bibinfo
    {pages} {4582} (\bibinfo {year} {2020})}\BibitemShut {NoStop}%
  \bibitem [{\citenamefont {Hayashida}\ \emph {et~al.}(2021)\citenamefont
    {Hayashida}, \citenamefont {Kimura}, \citenamefont {Urushihara},
    \citenamefont {Asaka},\ and\ \citenamefont {Kimura}}]{Hayashida2021}%
    \BibitemOpen
    \bibfield  {author} {\bibinfo {author} {\bibfnamefont {T.}~\bibnamefont
    {Hayashida}}, \bibinfo {author} {\bibfnamefont {K.}~\bibnamefont {Kimura}},
    \bibinfo {author} {\bibfnamefont {D.}~\bibnamefont {Urushihara}}, \bibinfo
    {author} {\bibfnamefont {T.}~\bibnamefont {Asaka}},\ and\ \bibinfo {author}
    {\bibfnamefont {T.}~\bibnamefont {Kimura}},\ }\href
    {https://doi.org/10.1021/jacs.1c00391} {\bibfield  {journal} {\bibinfo
    {journal} {J. Am. Chem. Soc.}\ }\textbf {\bibinfo {volume} {143}},\ \bibinfo
    {pages} {3638} (\bibinfo {year} {2021})}\BibitemShut {NoStop}%
  \bibitem [{\citenamefont {Zhang}\ and\ \citenamefont {Niu}(2015)}]{Zhang2015}%
    \BibitemOpen
    \bibfield  {author} {\bibinfo {author} {\bibfnamefont {L.}~\bibnamefont
    {Zhang}}\ and\ \bibinfo {author} {\bibfnamefont {Q.}~\bibnamefont {Niu}},\
    }\href {https://doi.org/10.1103/PhysRevLett.115.115502} {\bibfield  {journal}
    {\bibinfo  {journal} {Phys. Rev. Lett.}\ }\textbf {\bibinfo {volume} {115}},\
    \bibinfo {pages} {115502} (\bibinfo {year} {2015})}\BibitemShut {NoStop}%
  \bibitem [{\citenamefont {Chen}\ \emph {et~al.}(2018)\citenamefont {Chen},
    \citenamefont {Zhang}, \citenamefont {Niu},\ and\ \citenamefont
    {Zhang}}]{Chen2018}%
    \BibitemOpen
    \bibfield  {author} {\bibinfo {author} {\bibfnamefont {H.}~\bibnamefont
    {Chen}}, \bibinfo {author} {\bibfnamefont {W.}~\bibnamefont {Zhang}},
    \bibinfo {author} {\bibfnamefont {Q.}~\bibnamefont {Niu}},\ and\ \bibinfo
    {author} {\bibfnamefont {L.}~\bibnamefont {Zhang}},\ }\href
    {https://doi.org/10.1088/2053-1583/aaf292} {\bibfield  {journal} {\bibinfo
    {journal} {2D Mater.}\ }\textbf {\bibinfo {volume} {6}},\ \bibinfo {pages}
    {012002} (\bibinfo {year} {2018})}\BibitemShut {NoStop}%
  \bibitem [{\citenamefont {Zhu}\ \emph {et~al.}(2018)\citenamefont {Zhu},
    \citenamefont {Yi}, \citenamefont {Li}, \citenamefont {Xiao}, \citenamefont
    {Zhang}, \citenamefont {Yang}, \citenamefont {Kaindl}, \citenamefont {Li},
    \citenamefont {Wang},\ and\ \citenamefont {Zhang}}]{Zhu2018}%
    \BibitemOpen
    \bibfield  {author} {\bibinfo {author} {\bibfnamefont {H.}~\bibnamefont
    {Zhu}}, \bibinfo {author} {\bibfnamefont {J.}~\bibnamefont {Yi}}, \bibinfo
    {author} {\bibfnamefont {M.-Y.}\ \bibnamefont {Li}}, \bibinfo {author}
    {\bibfnamefont {J.}~\bibnamefont {Xiao}}, \bibinfo {author} {\bibfnamefont
    {L.}~\bibnamefont {Zhang}}, \bibinfo {author} {\bibfnamefont {C.-W.}\
    \bibnamefont {Yang}}, \bibinfo {author} {\bibfnamefont {R.~A.}\ \bibnamefont
    {Kaindl}}, \bibinfo {author} {\bibfnamefont {L.-J.}\ \bibnamefont {Li}},
    \bibinfo {author} {\bibfnamefont {Y.}~\bibnamefont {Wang}},\ and\ \bibinfo
    {author} {\bibfnamefont {X.}~\bibnamefont {Zhang}},\ }\href
    {https://doi.org/10.1126/science.aar2711} {\bibfield  {journal} {\bibinfo
    {journal} {Science}\ }\textbf {\bibinfo {volume} {359}},\ \bibinfo {pages}
    {579} (\bibinfo {year} {2018})}\BibitemShut {NoStop}%
  \bibitem [{\citenamefont {Kishine}\ \emph {et~al.}(2020)\citenamefont
    {Kishine}, \citenamefont {Ovchinnikov},\ and\ \citenamefont
    {Tereshchenko}}]{Kishine2020}%
    \BibitemOpen
    \bibfield  {author} {\bibinfo {author} {\bibfnamefont {J.}~\bibnamefont
    {Kishine}}, \bibinfo {author} {\bibfnamefont {A.~S.}\ \bibnamefont
    {Ovchinnikov}},\ and\ \bibinfo {author} {\bibfnamefont {A.~A.}\ \bibnamefont
    {Tereshchenko}},\ }\href {https://doi.org/10.1103/PhysRevLett.125.245302}
    {\bibfield  {journal} {\bibinfo  {journal} {Phys. Rev. Lett.}\ }\textbf
    {\bibinfo {volume} {125}},\ \bibinfo {pages} {245302} (\bibinfo {year}
    {2020})}\BibitemShut {NoStop}%
  \bibitem [{\citenamefont {Chen}\ \emph {et~al.}(2021)\citenamefont {Chen},
    \citenamefont {Kadic},\ and\ \citenamefont {Wegener}}]{Chen2021}%
    \BibitemOpen
    \bibfield  {author} {\bibinfo {author} {\bibfnamefont {Y.}~\bibnamefont
    {Chen}}, \bibinfo {author} {\bibfnamefont {M.}~\bibnamefont {Kadic}},\ and\
    \bibinfo {author} {\bibfnamefont {M.}~\bibnamefont {Wegener}},\ }\href
    {https://doi.org/10.1038/s41467-021-23574-2} {\bibfield  {journal} {\bibinfo
    {journal} {Nat. Commun.}\ }\textbf {\bibinfo {volume} {12}},\ \bibinfo
    {pages} {3278} (\bibinfo {year} {2021})}\BibitemShut {NoStop}%
  \bibitem [{\citenamefont {Ishito}\ \emph {et~al.}(2021)\citenamefont {Ishito},
    \citenamefont {Mao}, \citenamefont {Kousaka}, \citenamefont {Togawa},
    \citenamefont {Iwasaki}, \citenamefont {Zhang}, \citenamefont {Murakami},
    \citenamefont {ichiro Kishine},\ and\ \citenamefont
    {Satoh}}]{ishito2021truly}%
    \BibitemOpen
    \bibfield  {author} {\bibinfo {author} {\bibfnamefont {K.}~\bibnamefont
    {Ishito}}, \bibinfo {author} {\bibfnamefont {H.}~\bibnamefont {Mao}},
    \bibinfo {author} {\bibfnamefont {Y.}~\bibnamefont {Kousaka}}, \bibinfo
    {author} {\bibfnamefont {Y.}~\bibnamefont {Togawa}}, \bibinfo {author}
    {\bibfnamefont {S.}~\bibnamefont {Iwasaki}}, \bibinfo {author} {\bibfnamefont
    {T.}~\bibnamefont {Zhang}}, \bibinfo {author} {\bibfnamefont
    {S.}~\bibnamefont {Murakami}}, \bibinfo {author} {\bibfnamefont
    {J.}~\bibnamefont {ichiro Kishine}},\ and\ \bibinfo {author} {\bibfnamefont
    {T.}~\bibnamefont {Satoh}},\ }\href@noop {} {} (\bibinfo {year} {2021}),\
    \Eprint {https://arxiv.org/abs/2110.11604} {arXiv:2110.11604} \BibitemShut
    {NoStop}%
  \bibitem [{\citenamefont {Hamada}\ \emph {et~al.}(2018)\citenamefont {Hamada},
    \citenamefont {Minamitani}, \citenamefont {Hirayama},\ and\ \citenamefont
    {Murakami}}]{Hamada2018}%
    \BibitemOpen
    \bibfield  {author} {\bibinfo {author} {\bibfnamefont {M.}~\bibnamefont
    {Hamada}}, \bibinfo {author} {\bibfnamefont {E.}~\bibnamefont {Minamitani}},
    \bibinfo {author} {\bibfnamefont {M.}~\bibnamefont {Hirayama}},\ and\
    \bibinfo {author} {\bibfnamefont {S.}~\bibnamefont {Murakami}},\ }\href
    {https://doi.org/10.1103/PhysRevLett.121.175301} {\bibfield  {journal}
    {\bibinfo  {journal} {Phys. Rev. Lett.}\ }\textbf {\bibinfo {volume} {121}},\
    \bibinfo {pages} {175301} (\bibinfo {year} {2018})}\BibitemShut {NoStop}%
  \bibitem [{\citenamefont {Bouad}\ \emph {et~al.}(2003)\citenamefont {Bouad},
    \citenamefont {Chapon}, \citenamefont {Marin-Ayral}, \citenamefont
    {Bouree-Vigneron},\ and\ \citenamefont {Tedenac}}]{Bouad_2003}%
    \BibitemOpen
    \bibfield  {author} {\bibinfo {author} {\bibfnamefont {N.}~\bibnamefont
    {Bouad}}, \bibinfo {author} {\bibfnamefont {L.}~\bibnamefont {Chapon}},
    \bibinfo {author} {\bibfnamefont {R.-M.}\ \bibnamefont {Marin-Ayral}},
    \bibinfo {author} {\bibfnamefont {F.}~\bibnamefont {Bouree-Vigneron}},\ and\
    \bibinfo {author} {\bibfnamefont {J.-C.}\ \bibnamefont {Tedenac}},\ }\href
    {https://doi.org/https://doi.org/10.1016/S0022-4596(03)00017-3} {\bibfield
    {journal} {\bibinfo  {journal} {J. Solid State Chem.}\ }\textbf {\bibinfo
    {volume} {173}},\ \bibinfo {pages} {189} (\bibinfo {year}
    {2003})}\BibitemShut {NoStop}%
  \bibitem [{\citenamefont {Cheng}\ \emph {et~al.}(2019)\citenamefont {Cheng},
    \citenamefont {Wu}, \citenamefont {Zhu},\ and\ \citenamefont
    {Guo}}]{Cheng_2019}%
    \BibitemOpen
    \bibfield  {author} {\bibinfo {author} {\bibfnamefont {M.}~\bibnamefont
    {Cheng}}, \bibinfo {author} {\bibfnamefont {S.}~\bibnamefont {Wu}}, \bibinfo
    {author} {\bibfnamefont {Z.-Z.}\ \bibnamefont {Zhu}},\ and\ \bibinfo {author}
    {\bibfnamefont {G.-Y.}\ \bibnamefont {Guo}},\ }\href
    {https://doi.org/10.1103/PhysRevB.100.035202} {\bibfield  {journal} {\bibinfo
     {journal} {Phys. Rev. B}\ }\textbf {\bibinfo {volume} {100}},\ \bibinfo
    {pages} {035202} (\bibinfo {year} {2019})}\BibitemShut {NoStop}%
  \bibitem [{\citenamefont {Suzuki}\ \emph {et~al.}(2017)\citenamefont {Suzuki},
    \citenamefont {Koretsune}, \citenamefont {Ochi},\ and\ \citenamefont
    {Arita}}]{MTS_cmul_ahe_2017}%
    \BibitemOpen
    \bibfield  {author} {\bibinfo {author} {\bibfnamefont {M.-T.}\ \bibnamefont
    {Suzuki}}, \bibinfo {author} {\bibfnamefont {T.}~\bibnamefont {Koretsune}},
    \bibinfo {author} {\bibfnamefont {M.}~\bibnamefont {Ochi}},\ and\ \bibinfo
    {author} {\bibfnamefont {R.}~\bibnamefont {Arita}},\ }\href
    {https://doi.org/10.1103/PhysRevB.95.094406} {\bibfield  {journal} {\bibinfo
    {journal} {Phys. Rev. B}\ }\textbf {\bibinfo {volume} {95}},\ \bibinfo
    {pages} {094406} (\bibinfo {year} {2017})}\BibitemShut {NoStop}%
  \bibitem [{\citenamefont {Suzuki}\ \emph {et~al.}(2018)\citenamefont {Suzuki},
    \citenamefont {Ikeda},\ and\ \citenamefont {Oppeneer}}]{MTS_fp_magmul_2018}%
    \BibitemOpen
    \bibfield  {author} {\bibinfo {author} {\bibfnamefont {M.-T.}\ \bibnamefont
    {Suzuki}}, \bibinfo {author} {\bibfnamefont {H.}~\bibnamefont {Ikeda}},\ and\
    \bibinfo {author} {\bibfnamefont {P.~M.}\ \bibnamefont {Oppeneer}},\ }\href
    {https://doi.org/10.7566/JPSJ.87.041008} {\bibfield  {journal} {\bibinfo
    {journal} {J. Phys. Soc. Jpn.}\ }\textbf {\bibinfo {volume} {87}},\ \bibinfo
    {pages} {041008} (\bibinfo {year} {2018})}\BibitemShut {NoStop}%
  \bibitem [{\citenamefont {Hayami}\ \emph {et~al.}(2020)\citenamefont {Hayami},
    \citenamefont {Yanagi},\ and\ \citenamefont {Kusunose}}]{SH_bottom_up_2020}%
    \BibitemOpen
    \bibfield  {author} {\bibinfo {author} {\bibfnamefont {S.}~\bibnamefont
    {Hayami}}, \bibinfo {author} {\bibfnamefont {Y.}~\bibnamefont {Yanagi}},\
    and\ \bibinfo {author} {\bibfnamefont {H.}~\bibnamefont {Kusunose}},\ }\href
    {https://doi.org/10.1103/PhysRevB.102.144441} {\bibfield  {journal} {\bibinfo
     {journal} {Phys. Rev. B}\ }\textbf {\bibinfo {volume} {102}},\ \bibinfo
    {pages} {144441} (\bibinfo {year} {2020})}\BibitemShut {NoStop}%
  \bibitem [{\citenamefont {Oiwa}\ and\ \citenamefont
    {Kusunose}(2022)}]{ROHK_2022}%
    \BibitemOpen
    \bibfield  {author} {\bibinfo {author} {\bibfnamefont {R.}~\bibnamefont
    {Oiwa}}\ and\ \bibinfo {author} {\bibfnamefont {H.}~\bibnamefont
    {Kusunose}},\ }\href {https://doi.org/10.7566/JPSJ.91.014701} {\bibfield
    {journal} {\bibinfo  {journal} {J. Phys. Soc. Jpn.}\ }\textbf {\bibinfo
    {volume} {91}},\ \bibinfo {pages} {014701} (\bibinfo {year}
    {2022})}\BibitemShut {NoStop}%
  \bibitem [{sm_()}]{sm_comment}%
    \BibitemOpen
    \href@noop {} {}\bibinfo {note} {(Supplemental Material) The definition of
    the symmetry-adopted multipole basis and the explicit expressions of the TB
    Hamiltonian in terms of them are given in detail. A brief description of the
    process of parameter fitting, and the obtained values of the 30 parameters
    within the nearest-neighbor hopping are also given.}\BibitemShut {Stop}%
  \bibitem [{\citenamefont {Wang}\ \emph {et~al.}(2021)\citenamefont {Wang},
    \citenamefont {Ye}, \citenamefont {Wang}, \citenamefont {He}, \citenamefont
    {Huang},\ and\ \citenamefont {Chang}}]{Wang_2021}%
    \BibitemOpen
    \bibfield  {author} {\bibinfo {author} {\bibfnamefont {Z.}~\bibnamefont
    {Wang}}, \bibinfo {author} {\bibfnamefont {S.}~\bibnamefont {Ye}}, \bibinfo
    {author} {\bibfnamefont {H.}~\bibnamefont {Wang}}, \bibinfo {author}
    {\bibfnamefont {J.}~\bibnamefont {He}}, \bibinfo {author} {\bibfnamefont
    {Q.}~\bibnamefont {Huang}},\ and\ \bibinfo {author} {\bibfnamefont
    {S.}~\bibnamefont {Chang}},\ }\href
    {https://doi.org/10.1038/s41524-020-00490-5} {\bibfield  {journal} {\bibinfo
    {journal} {NPJ Comput. Mater.}\ }\textbf {\bibinfo {volume} {7}},\ \bibinfo
    {pages} {11} (\bibinfo {year} {2021})}\BibitemShut {NoStop}%
  \bibitem [{\citenamefont {Goto}\ \emph {et~al.}(1986)\citenamefont {Goto},
    \citenamefont {Tamaki}, \citenamefont {Fujimura},\ and\ \citenamefont
    {Unoki}}]{Goto_1986}%
    \BibitemOpen
    \bibfield  {author} {\bibinfo {author} {\bibfnamefont {T.}~\bibnamefont
    {Goto}}, \bibinfo {author} {\bibfnamefont {A.}~\bibnamefont {Tamaki}},
    \bibinfo {author} {\bibfnamefont {T.}~\bibnamefont {Fujimura}},\ and\
    \bibinfo {author} {\bibfnamefont {H.}~\bibnamefont {Unoki}},\ }\href
    {https://doi.org/10.1143/JPSJ.55.1613} {\bibfield  {journal} {\bibinfo
    {journal} {J. Phys. Soc. Jpn.}\ }\textbf {\bibinfo {volume} {55}},\ \bibinfo
    {pages} {1613} (\bibinfo {year} {1986})}\BibitemShut {NoStop}%
  \bibitem [{\citenamefont {Kuromaru}\ \emph {et~al.}(2001)\citenamefont
    {Kuromaru}, \citenamefont {Kusunose},\ and\ \citenamefont
    {Kuramoto}}]{Kuromaru_2000}%
    \BibitemOpen
    \bibfield  {author} {\bibinfo {author} {\bibfnamefont {T.}~\bibnamefont
    {Kuromaru}}, \bibinfo {author} {\bibfnamefont {H.}~\bibnamefont {Kusunose}},\
    and\ \bibinfo {author} {\bibfnamefont {Y.}~\bibnamefont {Kuramoto}},\ }\href
    {https://doi.org/10.1143/JPSJ.70.521} {\bibfield  {journal} {\bibinfo
    {journal} {J. Phys. Soc. Jpn.}\ }\textbf {\bibinfo {volume} {70}},\ \bibinfo
    {pages} {521} (\bibinfo {year} {2001})}\BibitemShut {NoStop}%
  \bibitem [{\citenamefont {Kurihara}\ \emph {et~al.}(2017)\citenamefont
    {Kurihara}, \citenamefont {Mitsumoto}, \citenamefont {Akatsu}, \citenamefont
    {Nemoto}, \citenamefont {Goto}, \citenamefont {Kobayashi},\ and\
    \citenamefont {Sato}}]{Kurihara_2017}%
    \BibitemOpen
    \bibfield  {author} {\bibinfo {author} {\bibfnamefont {R.}~\bibnamefont
    {Kurihara}}, \bibinfo {author} {\bibfnamefont {K.}~\bibnamefont {Mitsumoto}},
    \bibinfo {author} {\bibfnamefont {M.}~\bibnamefont {Akatsu}}, \bibinfo
    {author} {\bibfnamefont {Y.}~\bibnamefont {Nemoto}}, \bibinfo {author}
    {\bibfnamefont {T.}~\bibnamefont {Goto}}, \bibinfo {author} {\bibfnamefont
    {Y.}~\bibnamefont {Kobayashi}},\ and\ \bibinfo {author} {\bibfnamefont
    {M.}~\bibnamefont {Sato}},\ }\href {https://doi.org/10.7566/JPSJ.86.064706}
    {\bibfield  {journal} {\bibinfo  {journal} {J. Phys. Soc. Jpn.}\ }\textbf
    {\bibinfo {volume} {86}},\ \bibinfo {pages} {064706} (\bibinfo {year}
    {2017})}\BibitemShut {NoStop}%
  \bibitem [{\citenamefont {Proskurin}\ \emph {et~al.}(2017)\citenamefont
    {Proskurin}, \citenamefont {Ovchinnikov}, \citenamefont {Nosov},\ and\
    \citenamefont {Kishine}}]{proskurin_2017}%
    \BibitemOpen
    \bibfield  {author} {\bibinfo {author} {\bibfnamefont {I.}~\bibnamefont
    {Proskurin}}, \bibinfo {author} {\bibfnamefont {A.~S.}\ \bibnamefont
    {Ovchinnikov}}, \bibinfo {author} {\bibfnamefont {P.}~\bibnamefont {Nosov}},\
    and\ \bibinfo {author} {\bibfnamefont {J.}~\bibnamefont {Kishine}},\
    }\href@noop {} {\bibfield  {journal} {\bibinfo  {journal} {New J. Phys.}\
    }\textbf {\bibinfo {volume} {19}},\ \bibinfo {pages} {063021} (\bibinfo
    {year} {2017})}\BibitemShut {NoStop}%
  \end{thebibliography}

%apsrev4-2.bst 2019-01-14 (MD) hand-edited version of apsrev4-1.bst
%Control: key (0)
%Control: author (72) initials jnrlst
%Control: editor formatted (1) identically to author
%Control: production of article title (-1) disabled
%Control: page (0) single
%Control: year (1) truncated
%Control: production of eprint (0) enabled
%

\end{document}